# Where is the Edge of Chaos?

Ron Fulbright
University of South Carolina Upstate
Spartanburg, SC 29303

fulbrigh@uscupstate.edu

**ABSTRACT**
Previous study of cellular automata and random Boolean networks has shown emergent behavior occurring at the edge of chaos where the randomness (order/disorder) of internal connections is set to an intermediate critical value. The value at which maximal emergent behavior occurs has been observed to be inversely related to the total number of interconnected elements—the neighborhood size. However, different equations predict different values. This paper presents a study of one-dimensional cellular automata (1DCA) verifying the general relationship but finding a more precise correlation with the radius of the neighborhood rather than neighborhood size. Furthermore, the critical value of the emergent regime is observed to be very close to 1/e hinting at the discovery of a fundamental characteristic of emergent systems.

## 1. INTRODUCTION

Researchers in the fields of nonlinear dynamics and complexity theory study cellular automata and Boolean networks to investigate dynamic behavior of emergent systems. A cellular automaton is a grid of cells each occupying one of a finite set of states. The next state a cell occupies is determined by the states of a set of neighboring cells. In one-dimensional cellular automata (1DCA), the size of the neighborhood, $N$, is defined as the target cell plus $r$ neighboring cells on either side ($N = 2r + 1$).

In many ways similar to cellular automata, Boolean networks are a set of variables where the next value of each variable is dependent on a set of other variables it is connected to. The set of connected variables is similar to the neighborhood of cells in cellular automata.

Systems composed of interacting elements like cellular automata and Boolean networks exhibit emergent behavior (Bedeau, 1997; Goldstein, 1999; O'Connor, 1984). Emergent behavior features properties the individual elements themselves do not possess. Instead, these properties and behaviors emerge only when the elements interact in a wider whole. The canonical example of emergent behavior is a flock of birds. The behavior of a flock spontaneously arises from simple rules of interaction followed by individual birds and does not involve any central coordination. Thus, a flock exhibits many properties and characteristics not possessed by an individual bird such as the ability to flow around an obstacle.

The amount of freedom in the connections between elements is critical. Maximal emergent behavior has been found to emerge when the internal degree of disorder is tuned to an intermediate value ($\lambda_c$) between order and chaos—denoting a region called the *emergent regime* (Wolfram, 1984; 2002; Langton, 1986; Li, Packard, & Langton, 1990). This region of system behavior, called the *edge of chaos*, denotes a change in dynamics akin to a phase transition in physics in which the balance of creation and destruction of internal structures maximizes the system's ability to encode and process information.

Since the most robust behavior emerges at $\lambda_c$, engineers would like to know or be able to predict $\lambda_c$ for a system. Values of $\lambda_c$ have been determined for specific systems, however, no value of $\lambda_c$ generally applicable to all systems has been measured. Previous work indicates $\lambda_c$ is related in some way to the size of the neighborhood in cellular automata and the connectivity in Boolean networks, but these metrics predict different values of $\lambda_c$.

The research question is: can we verify either the cellular automaton metric or the Boolean network metric as correct? In this paper, we study the behavior of 1DCA by varying the degree of randomness and observing the effect on $\lambda_c$. Since it appears $\lambda_c$ is not predicable for a particular run, we hypothesize averaging $\lambda_c$ over several runs of the 1DCA will show agreement with either the cellular automaton metric or the Boolean network metric.

## 2. LITERATURE

Wolfram (1984; 2002) conducted an in-depth study of cellular automata and categorized system behavior into

four classes based on the nature of the emergent properties:

**Class I:** *Static Regime* – a unique homogeneous state having no dynamical behavior

**Class II:** *Ordered Regime* - simple stable structure(s) emerge sometimes exhibiting periodic behavior but do not evolve or propagate

**Class III:** *Chaotic Regime* – continual creation of extremely short-lived patterns with no persistence or coherent structure

**Class IV:** *Emergent Regime* - persistent evolving and propagating stable structures exhibiting long periodic behavior

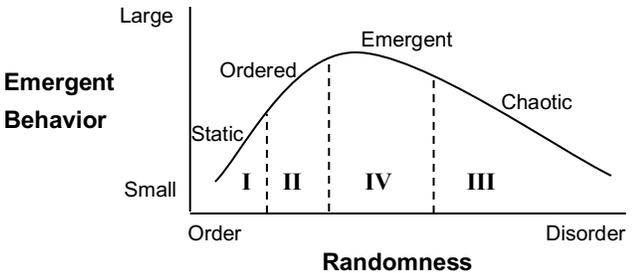

**Figure 1** – Wolfram identified four types of system behavior with Class IV corresponding to the emergent regime lying between the ordered and chaotic regimes and being dependent on the amount of disorder in the system. Reproduced from (Wolfram, 1984).

Figure 1 shows the location of the Wolfram classes with respect to the observed relationship between order, disorder, and emergent behavior. Similarly, Li, Packard & Langton (1990) identified six classes of behavior in cellular automata. Figure 2 compares the location of these classes to the Wolfram classes. Classes 2, 3, and 4 provide further detail of Wolfram Class II.

**[1]:** *Spatially homogeneous fixed points* – difference spreading rate, entropy, and mutual information is zero

**[2]:** *Spatially inhomogeneous fixed points* – fixed patterns, difference spreading rate is zero, entropy is finite

**[3]:** *Periodic behavior* – periodic behavior in isolated regions, entropy and mutual information is nonzero

**[4]:** *Locally chaotic behavior* – fixed or oscillating boundaries separating chaotic domains

**[5]:** *Chaotic behavior* – difference pattern spreading rate is high, mutual information is zero

**[6]:** *Complex behavior* - long transients, complex space time patterns, propagating structures, mutual information is large

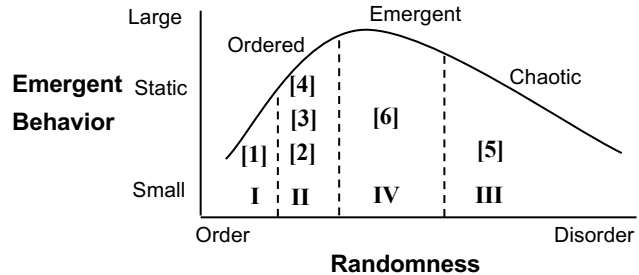

**Figure 2** – Li, Packard & Langton (1990) classes of system behavior (in brackets) compared to Wolfram (1984) classes.

Langton (1986) performed experiments with cellular automata varying the degree of randomness and observing behavior over time (subsequent generations of the cellular automata lattice). Langton called this the "lambda parameter" and identified $\lambda_c$ as the degree of randomness at which maximal dynamic behavior emerges as shown in Figure 3.

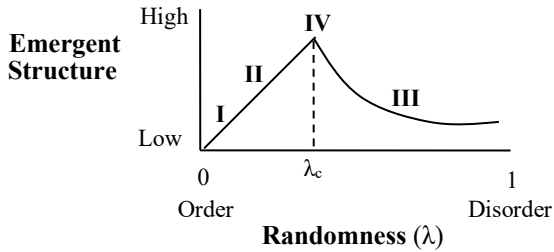

**Figure 3** – The $\lambda$ parameter allows one to change the amount of randomness (disorder) in the behavior of a 1-dimension cellular automaton. Maximal emergent behavior occurs near a critical value, $\lambda_c$, where system dynamics appears to undergo a phase change. The four Wolfram classes are shown in their approximate locations. Reproduced from (Langton, 1986).

Langton's $\lambda_c$ corresponds with Wolfram Class IV (between Wolfram classes II and III) and with Li, Packard & Langton's Class 6.

Highly ordered systems (Classes I, II, [1] [2] [3] [4]) are too rigid and tamp down dynamics so no emergent structure can arise. Highly disordered systems (Class III, [5]) destroy emergent structures so rapidly no patterns can form and persist. The emergent regime (Class IV, [6]) represents a balance between the formation and destruction of persistent structures. Such structures are necessary for the transmission, storage, and modification of information.

Li, Packard, & Langton (1990) found $\lambda_c$ to be dependent on the size of the neighborhood

$$\lambda_c \approx \frac{1}{2} - \frac{1}{2}\sqrt{1 - \frac{2}{r+1}}, \quad (1)$$

where $r$ is the radius of the neighborhood. For large values of $r$, this equation converges to

$$\lambda_c \approx \frac{1}{N}, \quad (2)$$

where $N$ is the size of the neighborhood ($2r+1$).

Researchers have studied other kinds of self-organizing networks. Kauffmann and others have used random Boolean networks (RBNs) to study phase transitions in dynamic behavior. In RBNs, the parameter $p_c$ controls the randomness of the interconnections between components (Kauffmann, 1993). This serves a similar role as the $\lambda_c$ parameter does in cellular automata. The critical value of $p_c$ is given by

$$p_c = \frac{1}{S}\left[1 + (1 - \frac{S}{K}[(2-S)K + S - 1])^{1/2}\right], \quad (3)$$

where $S$ is the number of states each node in the RBN can assume and $K$ is the number of neighbors to which the target node is connected. Note, $K$ for RBN is similar to $N$ for cellular automata.

Table 1 shows $p_c$ calculated for various values of $S$ and $K$ using Equation (3) for $p_c$ and Figure 3 shows $p_c$ plotted against $K$, the number of nodes connected together in a RBN, clearly indicating as $K$ increases, $p_c$ converges and decreases toward zero.

| K | S=2 | S=3 | S=4 | S=5 |
|---|-----|-----|-----|-----|
| 2 | 0.500 | 0.333 | 0.317 | 0.310 |
| 3 | 0.211 | 0.195 | 0.191 | 0.189 |
| 4 | 0.146 | 0.140 | 0.138 | 0.137 |
| 5 | 0.113 | 0.109 | 0.108 | 0.107 |
| 6 | 0.092 | 0.089 | 0.089 | 0.088 |
| 7 | 0.077 | 0.076 | 0.075 | 0.075 |
| 8 | 0.067 | 0.066 | 0.065 | 0.065 |
| 9 | 0.059 | 0.058 | 0.058 | 0.057 |
| 10 | 0.053 | 0.052 | 0.052 | 0.052 |

**Table 1** – The critical points, $p_c$, in random Boolean networks calculated for various number of states, $S$, and number of connected nodes, $K$.

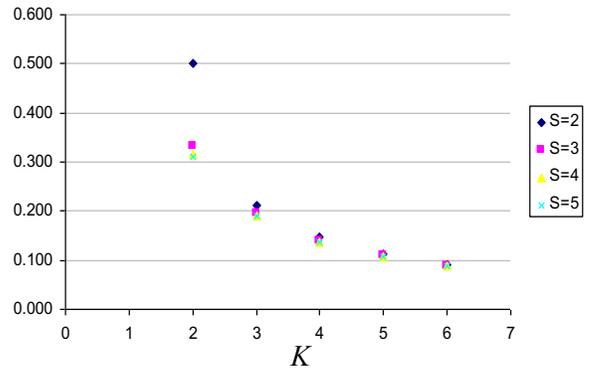

**Figure 3** – Plotting values from Table 1 shows a relationship between location of the emergent regime for RBNs, as indicated by $p_c$, and the size of the neighborhood, given by $K$.

There is good agreement in Equation (3) for all $K$ throughout the range of $K$ (except for very small $K$), so we can average the values of $p_c$ across all values of $K$ as shown in Table 2. Values of Equation (1) and Equation (3) for 1DCA are also shown in Table 2 for comparison.

Figure 4 compares Equation (1) and Equation (2) for 1DCA with Equation (3) for RBNs. For both CA and RBNs, the shape and form of the curves are consistent and clearly shows an inverse relationship between neighborhood size and the critical value of disorder in the system.

| N | r | $p_c$ Eq. (3) | $\lambda_c$ Eq. (1) | 1/N Eq. (2) |
|---|---|---|---|---|
| 2 | 0.5 | 0.365 | n/a | 0.500 |
| 3 | 1.0 | 0.197 | 0.500 | 0.333 |
| 4 | 1.5 | 0.140 | 0.276 | 0.250 |
| 5 | 2.0 | 0.109 | 0.211 | 0.200 |
| 6 | 2.5 | 0.090 | 0.173 | 0.167 |
| 7 | 3.0 | 0.076 | 0.146 | 0.143 |
| 8 | 3.5 | 0.066 | 0.127 | 0.125 |
| 9 | 4.0 | 0.058 | 0.113 | 0.111 |
| 10 | 4.5 | 0.052 | 0.101 | 0.100 |

**Table 2** – Averaging values of $p_c$ over several values of $S$ permits comparison of critical values for RBNs with critical values for 1DCA.

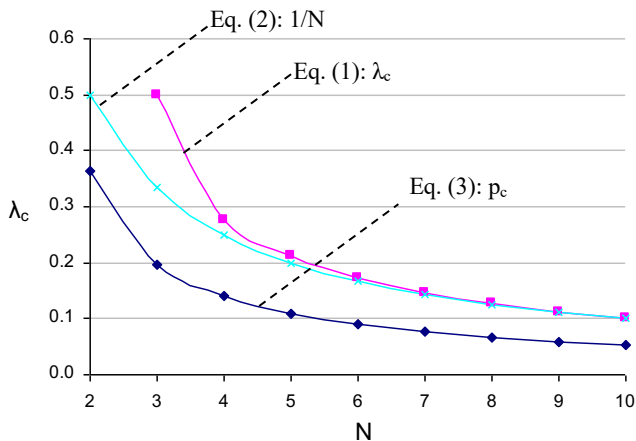

**Figure 4** – The critical value for disorder in a system is dependent on the size of the neighborhood. Equations (1) and (2) predict the critical value for 1DCA and Equation (3) predicts the critical value for RBNs.

Although Table 2 and Figure 4 demonstrate similar characteristics, they show the metric for determining $\lambda_c$ for CA gives different values than the metric for RBNs. The curve in Figure 4 for RBNs is offset downward from the curve for CA by a significant amount. Kauffman argues this difference is because RBNs have more degrees of freedom than CA. Specifically, the neighborhood in CA is defined to be the cells *physically* adjacent to the target cell. However, in RBNs, neighboring cells are not constrained to be physically nearby.

It is possible $\lambda_c$ is different for CA and RBN. Indeed, maybe a metric for $\lambda_c$ will have to be determined for every different kind of emergent system. However, we wonder if there is not a metric for $\lambda_c$ generally applicable for all systems. A first step toward this is to attempt to verify either the CA or RBN metric in an independent study.

## 3. METHODS

For this study we used a cellular automata applet developed by David Eck of the Hobart and William Smith Colleges in New York. The Java applet used in this study was found at:
http://math.hws.edu/xJava/CA/EdgeOfChaosCA.html
but is now available as a Javascript applet at:
https://math.hws.edu/eck/js/edge-of-chaos/CA.html

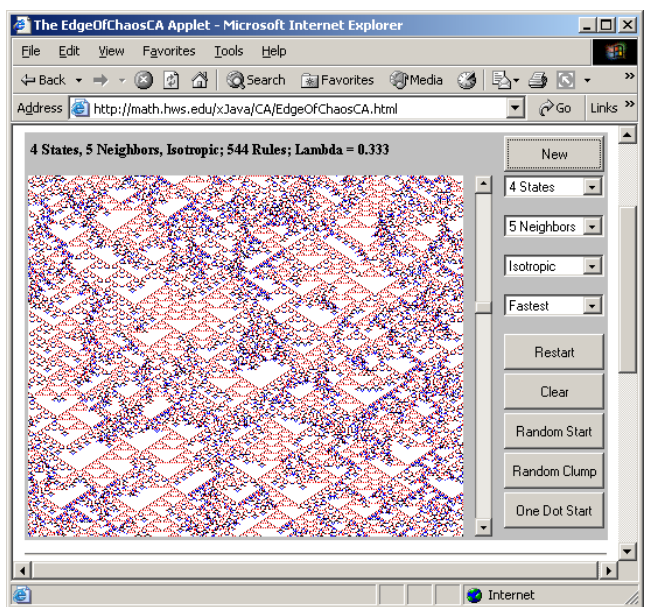

**Figure 5** – The cellular automata applet allows the user to change the number of states per cell, the number of neighbors, and the randomness (lambda). Generations of the cellular automata are displayed as rows from top to bottom on the screen.

Results were recorded with (states, neighborhood) set to: (3,3) (3,5) (3,7) (4,3) (4,5) (4,7) (5,3) (5,5) (6,3) (6,5) (7,3) and (7,5). For each setting, data were recorded for 25 runs. On each run, $\lambda$ was stepped through its entire range 0.0 to 1.0. The $\lambda$ value at which robust structure emerged was recorded. For runs exhibiting more than one emergent structure, the $\lambda$ value for each subsequent emergence was also recorded.

Figure 6 shows results of a typical run. Generally, no structure or simple steady-state structure was produced at lower values of $\lambda$ corresponding with Wolfram Class I behavior. At higher values of $\lambda$, more complex structure was produced but still exhibited steady-state characteristics—same non-evolving structure repeated every subsequent generation. This corresponds with Wolfram Class II behavior. At a critical value of λ, complex and evolving structure emerged. This is the value of λ recorded as λ_c for the run corresponding to Wolfram Call IV behavior. Emergent structures persisted throughout a range of λ but at higher values random noise took over corresponding to Wolfram Class III.

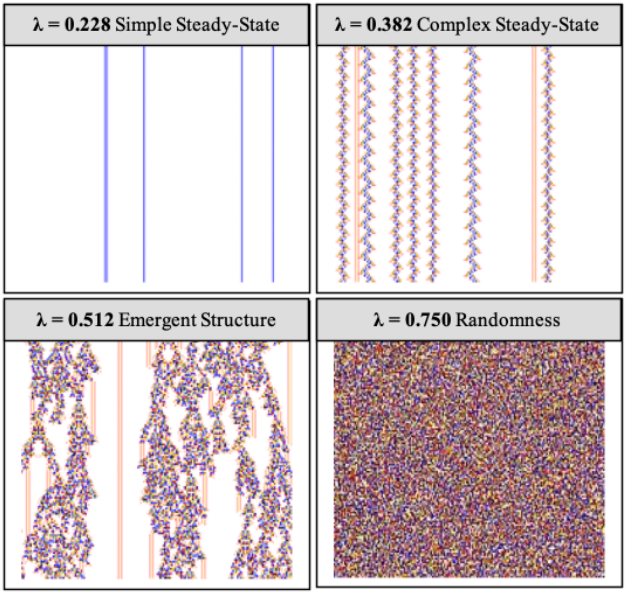

**Figure 6** – Dynamic behavior of IDCA changes as the degree of randomness, λ, changes from 0.0 (no randomness) to 1.0 (total randomness). Complex and evolving structure spontaneously emerges at a critical intermediate vale of λ.

Since every run exhibits a different output, 25 runs with the same setting of state and neighbor were made and results recorded. The emergent points were averaged over these runs.

## 4. RESULTS

Table 3 shows the average $\lambda_c$ for each value of states, $K$, and neighborhood size, $N$. These results are plotted in Figure 7. No particular pattern is observed when the values of observed $\lambda_c$ are plotted against $K$ but a very clear pattern emerges when plotted against $N$.

| K | N | λ |
|---|---|---|
| 3 | 3 | 0.646 |
| 4 | 3 | 0.510 |
| 5 | 3 | 0.488 |
| 6 | 3 | 0.526 |
| 7 | 3 | 0.441 |
| 3 | 5 | 0.284 |
| 4 | 5 | 0.331 |
| 5 | 5 | 0.295 |
| 6 | 5 | 0.301 |
| 7 | 5 | 0.299 |
| 3 | 7 | 0.259 |
| 4 | 7 | 0.261 |
| | | **0.387** |

**Table 3** – Average values of $\lambda_c$, at which emergent behavior is observed, for different number of states per cell (K) and different neighborhood size (N).

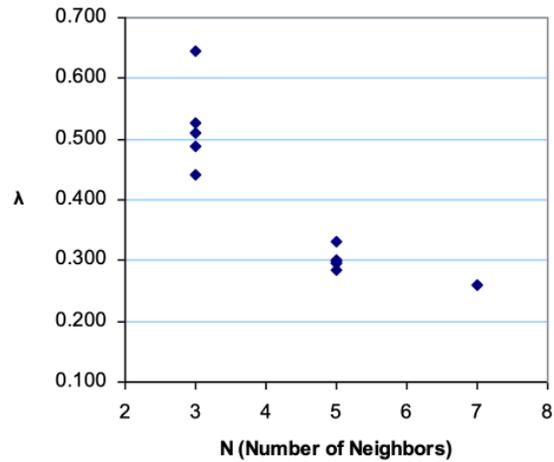

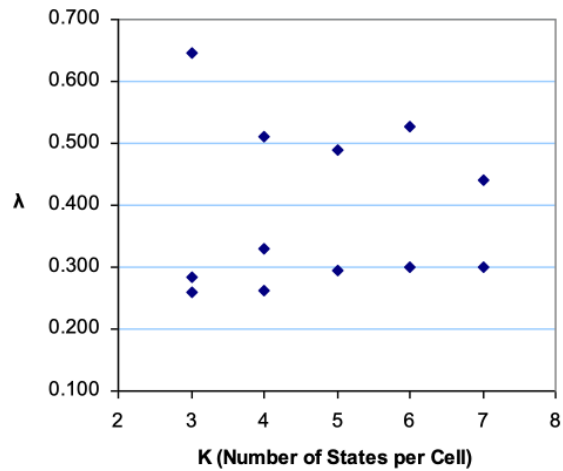

**Figure 7** – The observed values of $\lambda_c$, at which emergent behavior is observed, show a dependence on neighborhood size, N

The pattern observed in Figure 7 is similar to the pattern shown in Figure 4 for the equations of $\lambda_c$ for cellular automata and random Boolean networks. The shape of the curve is nearly identical, showing a decrease in $\lambda_c$ with increasing neighborhood size, $N$. Figure 8 compares our results to Equation (1) and (2) for cellular automata and Equation (3) from Boolean networks.

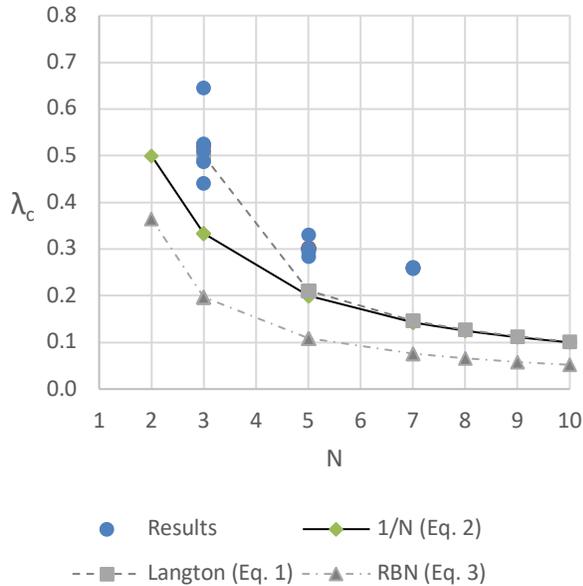

**Figure 8** – The observed values of $\lambda_c$, at which emergent behavior is observed, compared to equations from previous work on cellular automata (Eq. 1, 2) and random Boolean networks (Eq. 3).

However, our results lie significantly above either Equation (1), (2), or (3). We also notice our results are better characterized as being related to the radius of the neighborhood, $r$, instead of the size of the neighborhood, $N$. Recall, for 1DCA, $N = 2r + 1$. The best equation fitting our results is the simple relationship $1/(r +1)$ as shown in Figure 9.

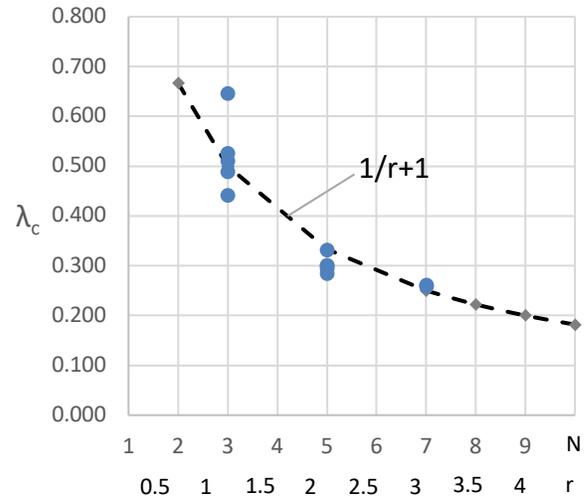

**Figure 9** – The observed values of $\lambda_c$, are related to the radius of the neighborhood, $r$.

## 5. CONCLUSION/FUTURE WORK

Refuted is our hypothesis averaging $\lambda_c$ over several runs of the 1DCA will show agreement with and confirm either the Li, Packard & Langton (1990) cellular automaton metric (Eq. 1, 2) or the Kauffmann (1993) random Boolean network equation (Eq. 3). As shown in Figure 8, our results, averaging $\lambda_c$ over 300 runs of the 1DCA Eck applet at various settings for the number of states per cell, $K$, and the size of the neighborhood, $N$, do not agree with either equation although the general shape of the curves agree. Therefore, we can verify the overall idea of $\lambda_c$ having an inverse relationship with the number of neighboring cells, but we find the relationship $1/(r +1)$ is a better fit for our results than either Eq. (1), (2), or (3). Thus, we propose

$$\lambda_c \approx \frac{1}{r+1}, \qquad (4)$$

where the neighborhood includes $r$ cells on each side of the target cell, as the equation predicting the amount of randomness at which emergent behavior emerges— the edge of chaos.

Whether or not Eq. (4) represents a general statement for all emergent systems is still an open question. It is possible our results are applicable only to the Eck applet which we used in this study. It is also possible the manner in which we obtained our results could have skewed the results. However, in our study, we took a statistical approach by averaging several runs, and then averaging the averages of these runs for different neighborhood sizes. We also observed emergent behavior directly rather than inferring emergent behavior by observing secondary measures such a mutual information as has been done in previous work. Therefore, we believe our results may indeed indicate a general truth for all emergent systems. Future work will have to investigate and verify this belief. Statistical studies of emergent systems other than 1DCA should be performed. Also, results by other researchers working with various types of emergent systems should be compared and contrasted with this study.

We also note, as shown in Table 3, the overall average of $\lambda_c$ 0.387 is very close (within 5%) to $1/e$ (0.368). The constant $e$ shows up in numerous applications across mathematics, science, and engineering. It would not be surprising if we eventually discover $e$ is involved in describing emergent behavior. We posit the following conjecture:

**The Emergent Regime 1/e Conjecture:** *A system consisting of multiple locally interacting elements will enter the emergent regime and exhibit global emergent behavior when the amount of randomness governing the interconnections between the elements ($\lambda_c$) is tuned to an intermediate value centering on 1/e (0.368, or about 37% of the range from no randomness to total randomness).*

We take $\lambda_c = 1/e$ to be a characteristic measure of the emergent regime. We do not anticipate $\lambda_c$ to be exactly 1/e for any particular emergent system, although this is certainly possible. Instead, we view this as a statistical measure. As an analogy, consider the batting average of a baseball player. Batting average is calculated by dividing the number of hits by the number of times at bat and ranges from 0–1. The average batting average of all major league baseball players is about 0.248. While some players may have a personal batting average of 0.248, this statistic does not say *all* baseball players will have or must have a batting average of 0.248. In fact, at any given point in time, *most* baseball players will have a batting average of some value other than 0.248. Instead of predicting performance of the individual, the batting average characteristic offers important insight for baseball by indicating the nature of the game, particularly the relationship between pitchers and hitters. Batters are successful only about 25% of the time. 75% of the time, batters do not get on base or get on base by means other than getting a hit.

Likewise, the 1/r conjecture informs engineers and system designers to tune the randomness in their systems to a range centering on 1/e for the most robust performance. Future work should compare the emergent points of other systems for fit to the 1/e conjecture.